\documentclass[twocolumn]{aastex61}

\newcommand{\MBD}{\Delta\tau_\mathrm{m}}
\newcommand{\SBD}{\Delta\tau_\mathrm{s}}

\usepackage{ulem}
\usepackage{color}
\usepackage{seqsplit}
\usepackage{hyperref}

\definecolor{blue}{rgb}{0.0,0.0,1.0}

\newcommand{\shao}{{Shanghai Astronomical Observatory, Chinese Academy of     
Sciences, Shanghai 200030, China}}

\submitjournal{AJ}

\begin{document}


\title{A Fast Radio Burst Search Method for VLBI Observation}
\author{Lei Liu}
\affiliation{\shao}

\author{Fengxian Tong}
\affiliation{\shao}

\author{Weimin Zheng}
\affiliation{\shao}
\affiliation{Key Laboratory of Radio Astronomy, Chinese Academy of 
	Sciences, Nanjing 210008, China}
\affiliation{Shanghai Key Laboratory of Space Navigation and Positioning 	
	Techniques, Shanghai 200030, China}

\author{Juan Zhang}
\affiliation{\shao}

\author{Li Tong}
\affiliation{\shao}

\correspondingauthor{Lei Liu}
\email{liulei@shao.ac.cn}

\begin{abstract}
We introduce the cross spectrum based FRB (Fast Radio Burst) search 
method for VLBI observation. This method optimizes the fringe fitting scheme in 
geodetic VLBI data post processing, which fully utilizes the cross spectrum 
fringe phase information and therefore maximizes the power of single pulse 
signals.
Working with cross spectrum greatly reduces the effect of radio 
frequency interference (RFI) compared with using auto spectrum. Single 
pulse detection confidence increases by cross 
identifying detections from multiple baselines. By combining the power of 
multiple baselines, we may improve the detection sensitivity. Our method is 
similar to that of coherent beam forming, but without the computational expense 
to form a great number of beams to cover the whole field of view of our 
telescopes.
The data processing pipeline designed for this method is easy to 
implement and parallelize, which can be deployed in various kinds of VLBI 
observations. In particular, we point out that VGOS observations are very 
suitable for FRB search.

\end{abstract}

\keywords{techniques: interferometric --- radio continuum: general --- 
	methods:data analysis --- pulsars: general}

\section{Introduction} \label{sec:intro}

Fast radio bursts (FRBs) are high flux radio flashes with milliseconds 
duration. At present the burst mechanism is still not  
clear \citep{Thornton2013, Keane2016, Chatterjee2017}. The high 
dispersion measure (DM) of current detections suggest their 
extragalactic origin . 
However, the possibility of Galactic origin still cannot be excluded, 
because 
the extra DM may be intrinsic to the source \citep{Katz2016a}. 

FRB was first discovered in 2007 
when reanalyzing the archive data of Parkes radio telescope 
\citep{Lorimer2007}. Until now, about 20 
FRBs are detected with large single dish telescopes and the 
UTMOST \citep{Caleb2016} interferometer. The angular resolution of these 
instruments is still not sufficient to unambiguously associate the burst with 
the background counterpart. Among those events, most of them are non-repeating 
burst. The only repeating burst FRB 121102 \citep{Spitler2016, Scholz2016} 
provides a good chance to obtain its high precision location via VLBI 
observation. 
\citet{Chatterjee2017} reveal that FRB 121102 originates within 100 mas of a 
faint 180 mJy persistent and compact radio source which is associate with an 
optical counterpart. \citet{Marcote2017} show for the first time that the 
bursts and the source are co-located with an angular separation of less than 12 
mas (a projected linear separation of less than 40~pc) by using the EVN 
(European VLBI Network) data. \citet{Tendulkar2017} classify the counterpart as 
a low-metallicity, star forming dwarf galaxy at a redshift of $z$ = 0.19273(8). 
\citet{Bassa2017} find the burst is coincident with a star forming region of 
the host galaxy. \citet{Scholz2017} carry out observations of the burst 
simultaneously in X-ray, Gamma-ray and radio band. 

Until now, it is still 
not clear if non-repeating and repeating bursts have the same origin. 
Catastrophic event models are able to explain non-repeating burst 
\citep{Dai2016}, including the merger of compact 
objects \citep{Kashiyama2013, Wang2016}, super massive neutron stars collapsing 
to black holes \citep{Falcke2014, Zhang2014}, and so on. The 
repeating burst is more likely to originate from young remnants of stellar 
collapses, neutron stars or black holes \citep{Katz2016a}, e.g. soft gamma 
repeaters \citep{Pen2015, Katz2016c}, giant pulse 
from young pulsars \citep{Keane2012, Katz2016b}, and the interaction of pulsars 
with planets \citep{Mottez2014}, asteroids or comets \citep{Geng2015, Dai2016}.

Nowadays FRB search is a hot topic in radio astronomy. The high 
precision localization of the burst is the key to identify background 
counterpart and finally explain the burst mechanism \citep{Masui2015} . 
Since the angular 
resolution of single dish telescope is low, it is very necessary to develop 
interferometer based method to achieve a higher resolution.
UTMOST \citep{Caleb2016} is the first radio interferometer search at 843 
MHz for fast transits, particularly FRB. Until 
2017, 3 FRBs have been discovered in a 180-day survey of the Southern sky 
\citep{Caleb2017}. 
However, its angular resolution (15~sec $\times$ 8.4~degree) is still not  
enough to identify background counterparts unambiguously. In this condition, 
VLBI \citep[Very Long Baseline Interferometer,][]{Whitney1976, 
Rogers1983, Thompson2001} search  with even higher angular 
resolution is very necessary.

There have been efforts to search for FRB signals in regular VLBI 
observations. The V-FASTR project \citep{VFASTR2011, Thompson2011} ran
commensally on VLBA and aimed at providing a few tens of 
milliarcsecond localization of detected FRB. Until 2016, the project had not 
detected any FRB, but it helped setting the upper limit of FRB event rate 
\citep{VFASTR2016}. The LOCATe project \citep{LOCATe} aimed at establishing an 
FRB pipeline for the EVN (European VLBI Network).
The project was based on the well-developed calibration pipeline and the EVN 
software correlator SFXC \citep{SFXC}. In the test phase, the pipeline had
successfully detected 
single pulses from test source RRAT J1819-1458 \citep{RRAT}, and localized the 
pulse position in image plane. In general, the pipeline of V-FASTR and LOCATe 
are similar.
We summarize them as the auto spectrum based method: They take the auto 
spectrum from the correlator, carry out RFI excising and 
dedispersion. After that standard pulsar search tools, e.g. 
PRESTO\footnote{
Available from the PRESTO website at 
\href{http://www.cv.nrao.edu/~sransom/presto}
{\seqsplit{http://www.cv.nrao.edu/\textasciitilde{}sransom/presto}}}
\citep{PRESTO} are used to detect single pulses. When a single pulse is 
identified in multiple stations, raw data at that time range are re-correlated, 
calibrated and finally used for localization.

VLBI technology has been widely used in the field of astrophysics 
\citep{Fish2016}, geodesy and astrometry \citep{Ma1998, Ma2009, Schuh2012, 
Behrend2013, ITRF}, deep 
space exploration \citep{Duev2012}. For geodetic VLBI, observations are 
scheduled in high frequency (e.g., once per week), and usually involve tens of 
stations for one session. Since large volume storages are very expensive, in 
standard data processing procedure, the large amount of VLBI raw observation 
data are deleted immediately after correlation. 
In authors' view, these data are not only important for geodetic
purpose, but also valuable for various kinds of time domain astronomy studies. 
It is such a good chance to carry out FRB search that the deletion of raw data 
is a huge loss. However, when developing our own FRB search pipeline for these 
data, we 
realize that the electromagnetic environment of some geodetic VLBI stations are 
not ideal. According to our study, the popular auto spectrum based method might 
not always behave well in single pulses extraction from RFI contaminated 
data\footnote{The 
testing data used in this work are taken from pulsar observation 
of geodetic VLBI stations. Data from Sh station are heavily contaminated by 
RFIs.}.
To fully exploit these data, we have to develop new method. 

In this work, we introduce the cross spectrum based single pulse search
method. This method takes the idea of geodetic VLBI data processing and fully 
utilizes the fringe phase information of cross spectrum. By carrying out fringe 
fitting on each cross spectrum of milliseconds duration, the power
of single pulse signal is maximized. In this way single pulses are extracted 
from given time range and baseline. 
Our method is good at extracting single pulses from RFI contaminated data,
which is very suitable for FRB search in various kinds of VLBI 
observations. 

This paper is organized as follows. In Sec. \ref{sec:method}, we introduce the 
cross spectrum based FRB search method, including the fringe fitting scheme, 
dedispersion and time segment construction, single pulse detection strategy, 
etc. In Sec. \ref{sec:cross}, we carry out single pulse search in a data set of 
VLBI pulsar observation and validate the search result with predicted pulsar 
phase. In Sec. \ref{sec:vgos}, we investigate the 
possibility of FRB search in VGOS observation. In Sec. \ref{sec:discuss}, we 
discuss issues not covered in previous sections. In Sec. \ref{sec:sum}, we 
summarize 
the whole work. 

\section{The cross spectrum based single pulse search method} 
\label{sec:method}
In this section, we introduce the cross spectrum based single 
pulse search method: VLBI raw data from each station are first correlated 
and output as APs (Accumulation period, or integration period) of milliseconds 
duration. Then several APs are summed 
together along time axis to construct time segments with given window length. 
After that fringe fitting is carried out independently for these time segments 
of multiple window lengths. Finally single pulse are extracted from each 
baseline and cross matched on multiple baselines. 

Besides above mentioned procedures, we have to point out that the search 
of dispersion measure is also an important part in the whole search
procedure. We have 
proposed a whole set of DM search scheme and present it in Sec. 
\ref{sec:dm_search}. For the pulsar observation data used in this work, the DM 
value is too small, which is not 
suitable for the testing of the DM search scheme. 

\subsection{VLBI correlation}

VLBI correlation is the fist step of the whole single pulse search procedure. 
In general, any FX type correlator that supports Mark5b \citep{Mark5b} or VDIF 
\citep{VDIF} format raw 
data decoding and is able to generate visibility (cross spectrum) output with 
milliseconds duration is competent for the work. Before the correlation of  
target scans, the clock offset and rate must be well adjusted 
with the calibration source, such that the residual delay is limited to one 
sample period and the fringe rate (residual delay rate multiplied with sky 
frequency) is no larger than $10^{-2}$~Hz. Since VLBI 
correlation is a time consuming process, a correlator that is able to fully 
exploit the computation power of modern CPU/GPU cluster is preferred. This is 
especially important for real time data correlation.

\subsection{Dedispersion and the construction of time segment} 
\label{sec:time_seg}
Let $S(n, j, k)$ represent the complex visibility of cross spectrum for a 
given baseline. Here $n,~j,~k$ correspond to the $n$-th sky frequency channel, 
the $j$-th frequency point inside the sky frequency channel and the $k$-th AP.
Due to dispersion, the arrival time of a single pulse is a function of 
dispersion measure (DM) and frequency. With given DM and reference frequency 
$f_\mathrm{ref, dm}$, the time offset $\Delta t$ of $j$-th frequency point 
in $n$-th frequency channel with respect to $f_\mathrm{ref,dm}$ is 
\citep{hpa}:
\begin{equation}\label{eq:dt_dm}
\Delta t	=	4.15\times 10^6 \times\mathrm{DM}\times\left[\frac{1}{(f_0^n + 
f_j)^2} - \frac{1}{f_\mathrm{ref,dm}^2}\right],
\end{equation}
where $f_0^n$ is the sky frequency of $n$-th channel, $f_j$ is the baseband 
frequency of $j$-th frequency point, time and frequency take the 
unit of millisecond and MHz, respectively. In this work, $f_\mathrm{ref, 
dm}$ always takes the highest frequency of the band, such that 
$t_\mathrm{offset}$ is nonnegative. $\Delta t$ is further mapped 
to the offset of AP index $\Delta k$:
\begin{equation}\label{eq:dk_dm}
\Delta k =	\lfloor 0.5 + \Delta t / t_\mathrm{ap}\rfloor,
\end{equation}
where $t_\mathrm{ap}$ is the duration of AP. The time segment is constructed by 
summing the cross spectrum within the window along the time axis:
\begin{equation}\label{eq:time_seg}
S_{k_0, l}(n, j)	=	\sum_{k = k_0 + \Delta k}^{k_0 + 
\Delta k + l - 1} S(n, j, k),
\end{equation}
where $S_{k_0, l}$ represents the time segment starting with AP index 
$k_0$ and summing along the time axis with a window length $l$; $n$, $j$ 
represent the $n$-th sky frequency channel and the $j$-th frequency point 
inside the sky frequency channel. Time segment is the basic 
unit of this work. We will carry out fringe fitting independently for  
time segments of different window lengths. One may find that $\Delta k$ is a 
function of $n, j$, which means in a time segment each frequency point might 
have its own starting and ending AP index. In Sec. \ref{sec:fringe_fitting}, we 
will explain the time segment constructed in this way can be used for fringe 
fitting as long as the corresponding fringe rate (delay rate times sky 
frequency) is small.

\subsection{Fringe fitting}\label{sec:fringe_fitting}
\begin{figure}
\plotone{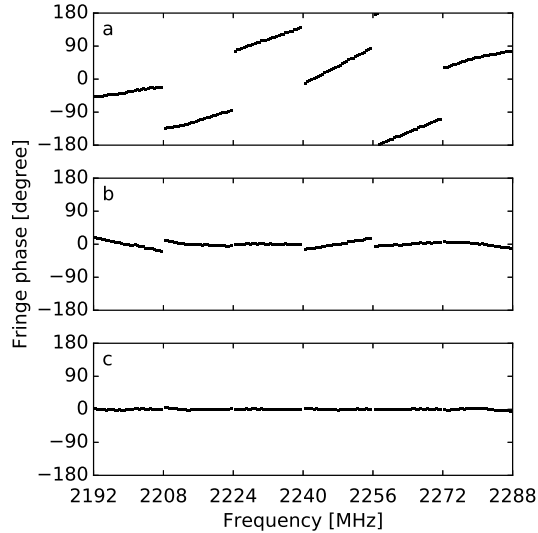}
\caption{Demonstration of fringe fitting process. Panel a: Fringe phase before 
fringe fitting. Panel b: Fringe phase after fringe fitting, with initial phase 
calibration only. Panel c: Fringe phase after fringe fitting, with both initial 
phase and channel delay calibration. Data are taken from calibration source 
3C273 of VLBI pulsar observation, Km-Ur baseline, 10~s duration. 6 stripes in 
each panel correspond to 6 frequency channels in S band. Each frequency channel 
contains 32 frequency points. The first point of each channel is not plotted in 
the figure, since it does not take part in fringe fitting as explained in Sec. 
\ref{sec:fringe_fitting}.\label{fig:fringe_fitting}}
\end{figure}
Fringe fitting is a key step in geodetic VLBI data post processing. As 
demonstrated in Fig. \ref{fig:fringe_fitting}, panel a, the fringe phase of 
visibility output is not flat before calibration, which leads to a low 
amplitude (power) of summed cross spectrum across the band. The purpose 
of fringe fitting is to find out residual delay, delay rate and initial phase 
across the band, and carry out fringe phase rotation accordingly 
\citep{Cotton1995, Takahashi2000, Cappallo2014}. After 
fringe fitting, the fringe phase across the whole band becomes flat and the 
amplitude of the summed cross spectrum is maximized. In this work, we take the 
algorithm of geodetic VLBI fringe fitting which is implemented in 
Haystack Observatory Postprocessing System (HOPS)\footnote{Available 
from the HOPS website at 
\href{https://www.haystack.mit.edu/tech/vlbi/hops.html}
{\seqsplit{https://www.haystack.mit.edu/tech/vlbi/hops.html}}},
 and make some modifications for higher accuracy and better performance.

The delay resolution function is defined as the summed cross 
spectrum after phase rotation \citep{Takahashi2000}:

\begin{eqnarray}
G(\SBD, \MBD, \Delta\dot{\tau}) = \sum_{n = 0}^{N-1} \sum_{j=1}^{J-1}  
\sum_{k=0}^{K-1} S(n, j, k)e^{-i\Phi(n, j, k)} \nonumber
\end{eqnarray}
with
\begin{eqnarray}\label{eq:fringe_fitting}
\Phi(n, j, k) & = & 2\pi f_j\SBD + 2\pi (f_0^n - f_\mathrm{ref,fit})\MBD 
\nonumber \\
			& + & 2\pi f_0^n\Delta\dot{\tau}t_\mathrm{ap} k + \Delta\phi_0^n, 
\end{eqnarray}
where $\SBD, \MBD, \Delta\dot{\tau}$ are the multi band residual delay (MBD), 
the single band residual delay (SBD) and the delay rate, respectively; 
$f_\mathrm{ref,fit}$ is the reference frequency for fringe fitting; 
$\Delta\phi_0^n$ is 
the initial phase of $n$-th frequency channel.  The triple 
summations are made for $N$ frequency channels across the band, $J-1$ 
frequency points inside one frequency channel and $K$ APs. Note that the 
summation of frequency points excludes $j=0$, so as to eliminate the offset of 
the correlation function \citep{Takahashi2000}. The fringe fitting procedure 
finds out MBD, SBD and delay rate that maximize the delay resolution function.

In a regular VLBI correlation, the fringe rate ($f_0^n\Delta\dot{\tau}$) of 
the cross spectrum is well limited to $10^{-2}$~Hz. The corresponding phase 
change due to fringe rate is small in a short time range (e.g. hundred 
milliseconds). As a modification to the standard fringe fitting algorithm, in 
this work, the term $2\pi f_0^n\Delta\dot{\tau}t_\mathrm{ap} k$ is neglected in 
Eq. \ref{eq:fringe_fitting}. A direct consequence is APs of 
milliseconds duration can be summed up directly with Eq. \ref{eq:time_seg} to 
construct a time segment. Fringe fitting now becomes SBD and MBD 
search for each time segment, which reduces computation complexity.

Eq. \ref{eq:fringe_fitting} is based on the assumption that all frequency 
channels share the same SBD. However, in real situation, channel delays exist 
due to the non-ideal instrument, which lead to the different slopes of fringe 
phase in each frequency channel. Since we use only one SBD to fit 
all frequency channels, the resulting fringe phase across the band are not 
flat (Fig. \ref{fig:fringe_fitting}, panel b). According to our study, this 
leads to the 5 - 10 percent loss of power
compared with the ideally flat fringe phase. For geodetic VLBI that always 
observe strong sources with hundred seconds duration, current fringe fitting 
scheme is good enough to achieve the desired SNR (signal to noise ratio). 
However, in this work, we detect single pulse signals with 
milliseconds duration. This extra power is very important. To make fringe phase 
flat, one direct approach is to fit SBD for every individual frequency channel. 
Obviously this is time consuming. We take another approach, by making an extra 
phase rotation $-2\pi(f_j - f_\mathrm{bw} / 2)\Delta\tau_0^n$ for every 
frequency point in a time segment before fringe fitting. Here $\Delta\tau_0^n$ 
comes from 
the calibration source, which is derived by fitting the channel delay of 
$n$-th frequency channel. After this phase rotation, all frequency channels of 
the 
time segment share the same SBD. The fringe phase across the band becomes 
much flatter after fringe fitting, which is shown in Fig. 
\ref{fig:fringe_fitting}, panel c. Note that above treatment is 
based on an assumption that delay differences between channels are constant 
in 
the whole observation. According to our fit to the 24 hours geodetic VLBI 
observation session, this is roughly the case. 

After above two modifications, Eq. \ref{eq:fringe_fitting} becomes:
\begin{eqnarray}
G_{k_0,~l}(\SBD, \MBD) = \sum_{n = 0}^{N-1} \sum_{j=1}^{J-1}  
S_{k_0,~l}(n, j) e^{-i\Phi(n, j)} \nonumber
\end{eqnarray}
with
\begin{eqnarray}\label{eq:fs_mod}
\Phi(n, j)	& = & 2\pi f_j\SBD + 2\pi (f_0^n - f_\mathrm{ref,fit})\MBD 
\nonumber \\  
			& + & 2\pi (f_j - f_\mathrm{bw} / 2)\Delta\tau_0^n + \Delta\phi_0^n,
\end{eqnarray}
where 
$G_{k_0,~l}(\SBD, \MBD)$ represents the delay resolution function of time 
segment $S_{k_0,~l}$. A relatively detailed explanation of channel 
delay $\Delta\tau_0^n$ and initial phase $\Delta\phi_0^n$ extraction is given 
in appendix \ref{append:pcal}. 

\begin{figure*}
	\fig{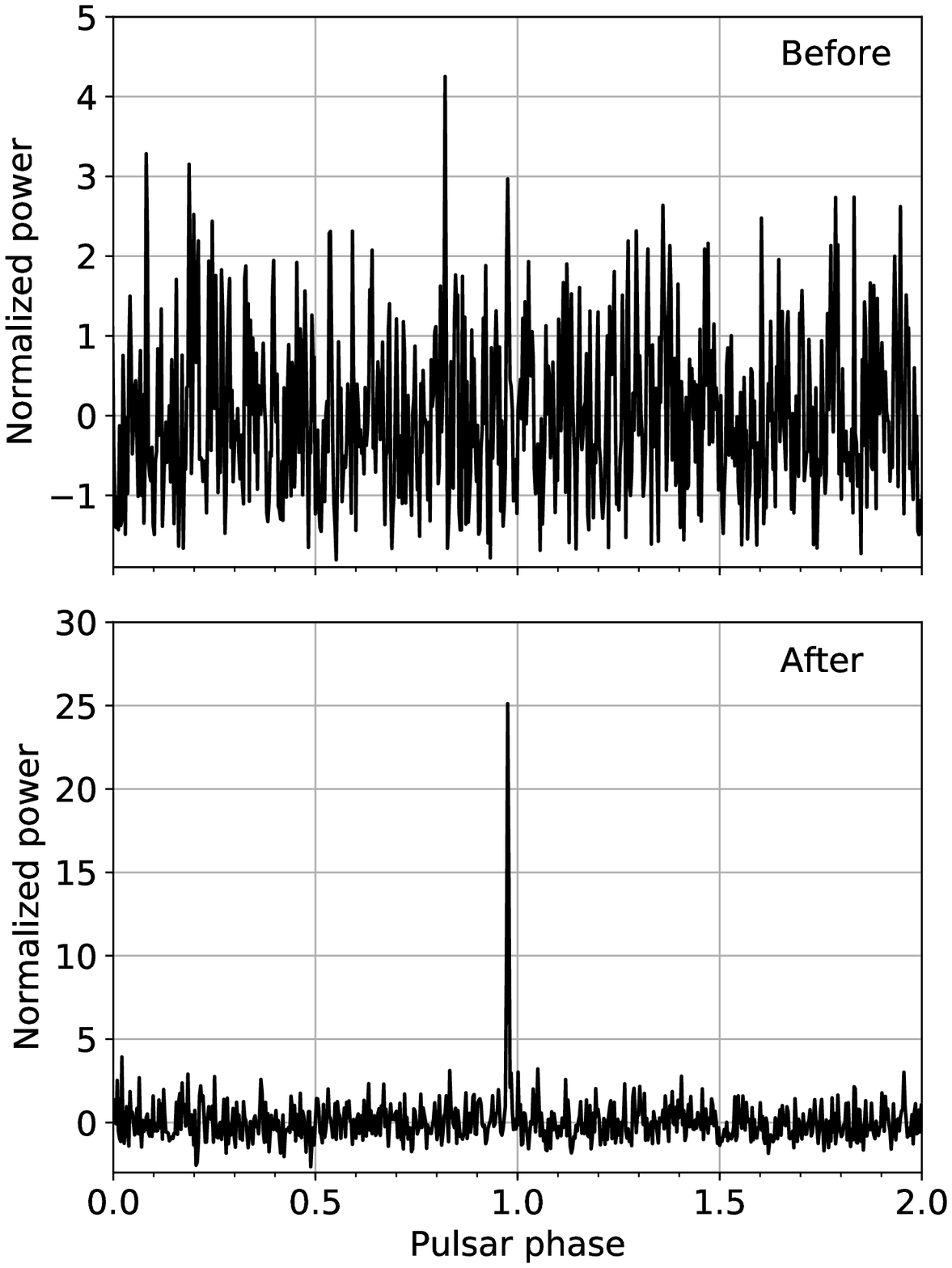}{0.45\textwidth}{(a)}
	\fig{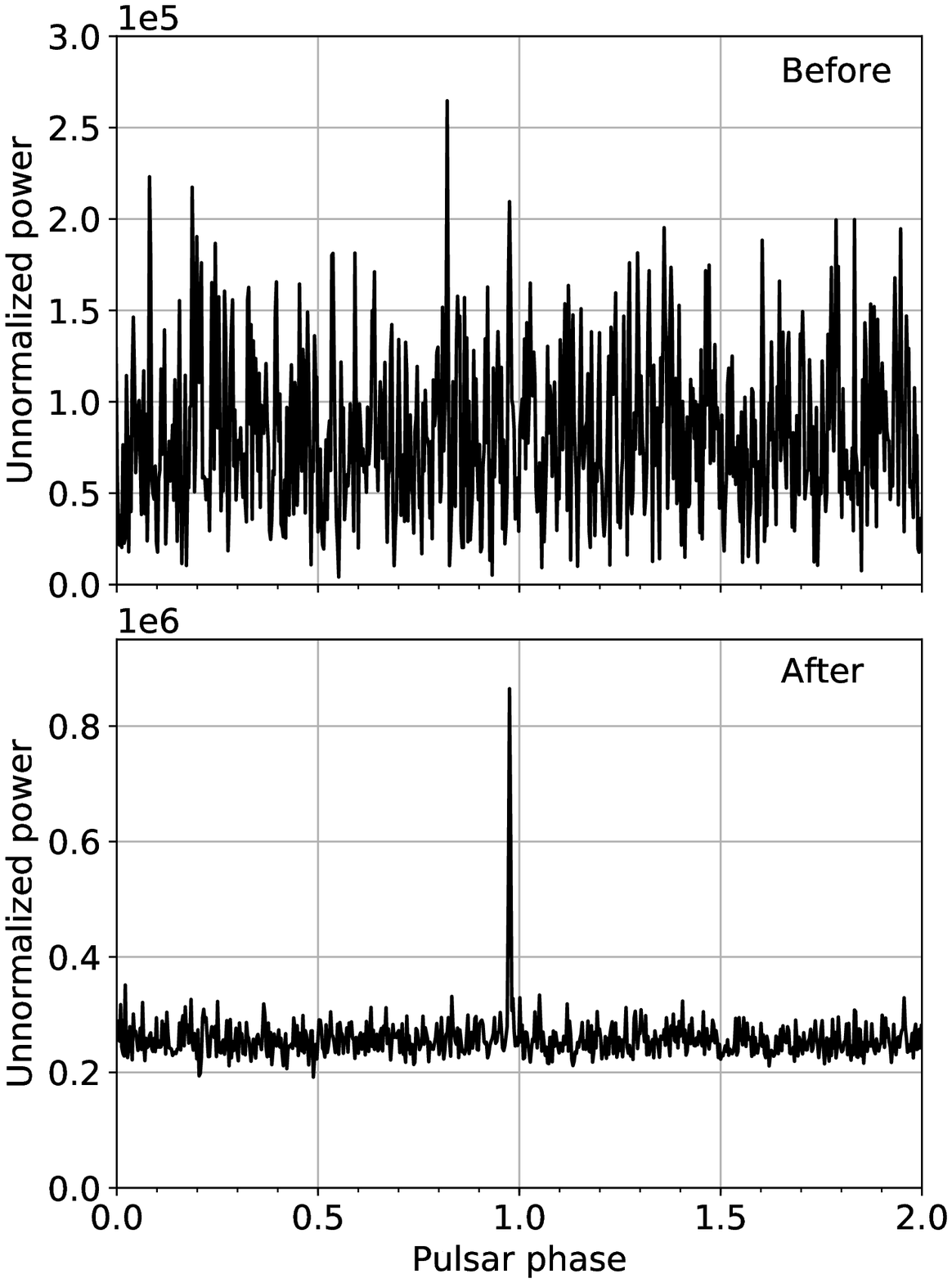}{0.45\textwidth}{(b)}
	\caption{Signal power before and after fringe fitting. The pulsar 
		phase and power of each time segment in the given time range is 
		plotted. Panel a and b present the normalized and unnormalized (raw) 
		power, 
		respectively. The single pulse is detected at 49.1~s of scan 73, Sh-Ur 
		baseline, and is shown as a cyan circle at corresponding time and scan 
		in Fig. 
		\ref{fig:result_bl}. The window length of time segments demonstrated in 
		this 
		figure is 4.096~ms. The phase of the single pulse is consistent with 
		the 
		pulsar phase range (0.973 - 0.983).\label{fig:fitdump}}
\end{figure*}
Panel a of Fig.~\ref{fig:fitdump} demonstrates the normalized power of each 
time segment before and after fringe fitting. It clearly shows that the 
normalized power of the single pulse around pulsar phase 0.978 is less than the 
threshold (we set it to 3 in Sec. \ref{sec:method_sp_extraction}) before fringe 
fitting, and is therefore not detectable. After fringe fitting, the 
power of the single pulse signal is greatly enhanced.
The normalized power is defined as:
\begin{equation}\label{eq:power_norm}
P_\mathrm{norm} = \left(|G| - |G|_\mathrm{average}\right)~/~|G|_\sigma,
\end{equation}
where $|G|$ is the amplitude of delay resolution function 
(summed power of cross spectrum) for each time 
segment,
$|G|_\mathrm{average}$ is the average of $|G|$ in a given time range,
$|G|_\sigma$ is the standard deviation of $|G|$.
Note the calculation of normalized power are baseline based.  
Panel b of Fig.~\ref{fig:fitdump} demonstrates the unnormalized power. 
As shown in the bottom panel of Fig.~\ref{fig:fitdump}.b, most of the 
time no single pulse signal is detected. The average power 
$|G|_\mathrm{average}$
corresponds to the system noise \citep{hpa}, and can be estimated with 
correlation theory:
\begin{equation}\label{eq:noise}
S_\mathrm{noise} = \sqrt{\frac{\mathrm{SEFD}_1\cdot\mathrm{SEFD}_2}{2BT}},
\end{equation}
where $S_\mathrm{noise}$ is in unit of Jansky (Jy), $B$ is the observation 
bandwidth, $T$ is the integration time, which corresponds to the window length 
in this work. The deviation of Eq.~\ref{eq:noise} takes the definition 
of SNR in \citet{Takahashi2000}, and assumes the normal observation conditions 
that the antenna temperature of the radio source is much smaller than the 
system noise.
 Generally speaking, 
it is not required that the signal power is much larger than the average noise 
as demonstrated in Fig. \ref{fig:fitdump}. A single pulse is already detectable 
if its power is several times larger than the noise fluctuation $|G|_\sigma$. 
This is the reason the detection threshold is in unit of noise fluctuation in 
this work.

\subsection{Single pulse extraction} \label{sec:method_sp_extraction}
We test two schemes of extracting single pulses from multiple baselines. In the 
first scheme, candidate signals are first detected and filtered on multiple 
windows, then they are cross matched on multiple baselines. In the second 
scheme, the power of multiple baselines are summed together to achieve a higher 
sensitivity, then single pulses are extracted via multiple windows filtering. 
Below we  describe these two schemes in detail.

\subsubsection{Multiple baselines cross matching scheme} 
\label{sec:method_cross_match}
\noindent\textbullet~~{Single pulse extraction from one baseline}\\
\begin{figure}
	\plotone{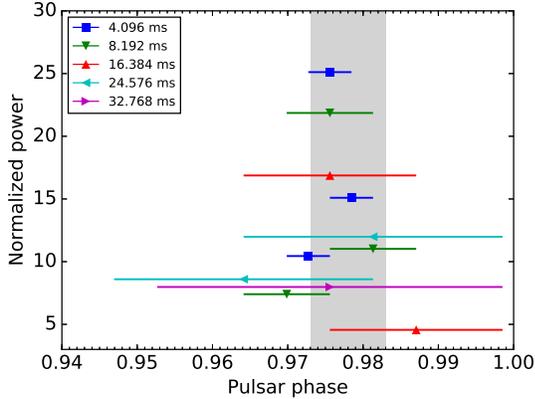}
	\caption{Single pulse detected in multiple windows. The single 
	pulse shown here is the same one as in Fig. \ref{fig:fitdump}. Error bars 
	of different length correspond to the range of time segments with their 
	respect window lengths. The gray region corresponds to the pulsar phase 
	range (0.973 - 0.983).\label{fig:pulse}} 
\end{figure}
For each baseline, we prepare several lists of time segments with different 
window lengths. The length of windows may range from one to ten times of a 
single pulse duration. By filtering signals with multiple 
windows, most RFI can be effectively excluded. 
Note that two consecutive time segments inside a list are half overlapped, e.g, 
the time segments are constructed as $S_{k_0, l}$, $S_{k_0+l/2, l}$, $S_{k_0+l, 
l}$, etc. This is to prevent the possible detection failure when a single pulse 
locates at the border of two interconnect time segments. 

After the construction of time segments, we carry out fringe fitting for each 
time segment and derive the normalized power with Eq. \ref{eq:power_norm}. 
In this work, we set a threshold of 3 for normalized power, signals with 
normalized power higher 
than 3 are kept for further filtering. For each window, a series of candidate 
signals are identified. Then 
the candidate signals are filtered with multiple windows to exclude RFI.
The reason to use multiple windows is based on the assumption: 
a single pulse with FRB or pulsar origin should be detectable in 
more than one window, and its MBD values in these windows are similar. In 
contrast, RFIs
with terrestrial origin would not be so stable and their MBD values are 
quite different in these windows. In general, the procedure of single pulse 
detection on one baseline is summarized below.
\begin{itemize}
	\item[a] Constructing time segments with multiple window lengths. Carrying 
	out fringe fittings for these time segments. Selecting candidate signals. A 
	selected candidate signal is uniquely described with time range, MBD and 
	normalized power. 
	\item[b] Starting with the largest window, inserting candidate signals of 
	each window to candidate groups. If the time range of a candidate 
	signal is overlapped with any candidate signal that belongs to a 
	candidate group, inserting it to that group; otherwise creating a new 
	candidate group and inserting again.
	\item[c] Calculating the average value of MBD $\bar{\MBD}$. Removing 
	candidate signals with MBD outside the range ($\bar{\MBD} - 
	t_\mathrm{ambig} 
	/ 4,~\bar{\MBD} + t_\mathrm{ambig} / 4$). Here $t_\mathrm{ambig}$ is the 
	ambiguity of MBD \citep{Takahashi2000} and must be take into 
	account when calculating average and removing data.
	\item[d] Calculating the number of different windows $N_\mathrm{win}$ in 
	the candidate group. A single pulse is assumed to be found if it is 
	detected in at least $N_\mathrm{win, min}$ windows in the group. The 
	candidate signal with the highest power is assumed to be the best 
	estimation of the single pulse, and is kept for multiple baselines cross 
	matching in the next step. The corresponding power and width are 
	regarded as the characteristic power and width of the single pulse.
\end{itemize}

Fig. \ref{fig:pulse} presents a single pulse detected on multiple windows. In 
some windows, several time segments catch the single pulse with different 
powers. It clearly shows that time segment of which the window length is 
comparable with the single pulse and the location is inside the pulsar phase 
range yields the highest power. Therefore, the cross spectrum method is able to 
estimate the width and location of the single pulse with the resolution of 
window length.\\

\noindent\textbullet~~{Cross matching of multiple baselines}

After single pulses are detected on multiple baselines, they are inserted to 
the cross matching groups. If the time range of one single pulse is overlapped 
with any single pulse in a cross matching group, it is inserted to that group; 
otherwise a new cross matching group is created and the single pulse is 
inserted. After insertion, we calculate the number of different baselines in 
each cross matching group. Single pulses detected on 2 or more baselines can 
almost excluded the possibility of RFI. \\

\subsubsection{Multiple baselines power summation scheme} 
\label{sec:method_power_sum}
In the test of single pulse detection from pulsar observation data, we 
find that the cross matching scheme is quite accurate in detecting  
pulsar signals. However, only a small number of pulsar signals are detected on 
low sensitivity baselines (e.g., Sh related baselines). The main reason is a 
significant amount of pulsar signals are so weak that they are not detectable 
if the baseline sensitivity is too low.
To improve the detection sensitivity, in this scheme, we first combine the 
power of several 
baselines, then extract single pulses via multiple windows filtering. 
The detailed implementation is described below.
\begin{itemize}
	\item[a] Preparing time segments of multiple window lengths for each 
	baseline and then carrying out fringe fitting for these time segments. This 
	step is the same as step a of previous scheme.
	\item[b] For each baseline, calculating the power (amplitude of delay 
	resolution function) of each time 
	segment after fringe fitting. Combining several baselines by summing the 
	power of corresponding time segments together, then selecting candidate 
	signals with given threshold. This is done for time segments of each window 
	length. 
	\item[c] Filtering the single pulse with multiple windows, which is the 
	same as step b in previous scheme. For the combined baseline, the MBD 
	information is of no use. Therefore step c of previous scheme is skipped.
	\item[d] Extracting single pulses from the candidate group, which is the 
	same as step d in previous scheme.
\end{itemize}
According to our test, we indeed improve the detection sensitivity for a 
less sensitive station by combining baselines of that station.

\section{Single pulse detection in VLBI pulsar observation} 
\label{sec:cross}
In this section, we exhibit our detection of single pulses from a data set of 
VLBI pulsar observation using the cross spectrum based method. Since the 
observed pulsar has been well studied, we could 
validate if a singe pulse originates from the pulsar or not via the predicted 
pulsar phase generated by TEMPO2 \citep{TEMPO2}.

\subsection{Observation overview}
\begin{deluxetable}{ll}
	\caption{Parameter setting of VLBI pulsar observation.\label{tab:obs}}
	\tablewidth{0pt}
	\tablehead{
	\colhead{Parameter} & \colhead{Setting}
	}
	\startdata
	Experiment code		&	psrf02 \\
	Observation date	&	Feb. 15, 2015  (MJD 57068) \\
	Observation time	&	Start:~UTC 08h00m00s \\
						&	Stop:~~UTC 20h01m40s \\
	Station				&	Sh, Km, Ur \\
	Target source		&	PSR J0332+5434 \\
	Reference source	&	J0347+5557 \\
	Calibration source	&	J1044+719, 3C273 \\
	Frequency band		&	S, X \\
 	Data collector		&	CDAS \citep{Zhu2016} \\
	Polarization 		&	Right circular \\
	Frequency channel	&	6 in S band, 10 in X band \\
	Bandwidth per channel	&	16~MHz \\
	Sample bits			&	2 \\
	\enddata
\end{deluxetable}
The VLBI data used in this work are taken from Chinese VLBI Network 
\citep[CVN,][]{CVN} pulsar observation in 2015 \citep{Chen2015}. Three 
CVN antennas (Sh\footnote{There are two telescopes in Shanghai: the 
Sheshan 
25~m and the Tianma 65~m. All tests in this work are based on Sheshan 
25~m data.}, Km, Ur) take part in the observation. Their performance are 
listed in Tab. \ref{tab:antenna}.
The main purpose of this observation is 
to obtain the accurate position of PSR J0332+5434 (B0329+54) using VLBI phase 
reference method \citep{Thompson2001}. PSR J0332+5434 is one of the brightest 
pulsars in the Northern sky. The average flux at S band is 0.1~Jy 
\citep{Kramer2003}. According to ATNF Pulsar 
Catalogue\footnote{\href{http://www.atnf.csiro.au/research/pulsar/psrcat}
{http://www.atnf.csiro.au/research/pulsar/psrcat}} 
\citep{ATNF}, the 
pulsar has a dispersion measure of 26.833 $\mathrm{pc~cm}^{-3}$ and a period of 
0.7145~s. Parameters of this observation are listed in Tab. 
\ref{tab:obs}. 

\begin{deluxetable}{ccc}
	\caption{Performance of 3 CVN 
	antennas.\footnote{\href{http://www.evlbi.org/user_guide/EVNstatus.txt}
	{http://www.evlbi.org/user\_guide/EVNstatus.txt}} \label{tab:antenna}}
	\tablewidth{0pt}
	\tablehead{
		\colhead{Antenna} & \colhead{Diameter} & \colhead{SEFD} \\
		\colhead{name}	&	\colhead{(m)}	&	\colhead{(Jy)}
	}
	\startdata
	Sh	&	25	&	800 \\
	~Km	&	40	&	350 \\
	Ur	&	25	&	560 \\
	\enddata
\end{deluxetable}

In this work, we use 6 frequency channels in S band for single pulse search. 
The 96~MHz bandwidth in S band (2192~MHz - 2288MHz) is divided into 6 16~MHz 
interconnected frequency channels\footnote{In geodetic VLBI observations, it is 
not necessary for frequency channels to be interconnected with each other.}. 
Each frequency channel 
contains 32 frequency points. As explained in Sec. \ref{sec:fringe_fitting}, 
the first frequency point of each channel is excluded for fringe fitting.

According to theoretical calculation \citep{Takahashi2000}, Km-Ur and Sh-Ur 
baseline yield the highest and lowest SNR, respectively. Besides 
that, the data receiving and collecting system of Sh antenna are influenced by 
low frequency electromagnetic interference, which further reduces 
its sensitivity. At present Sh station is mainly used for VLBI observations.

\begin{figure}
	\plotone{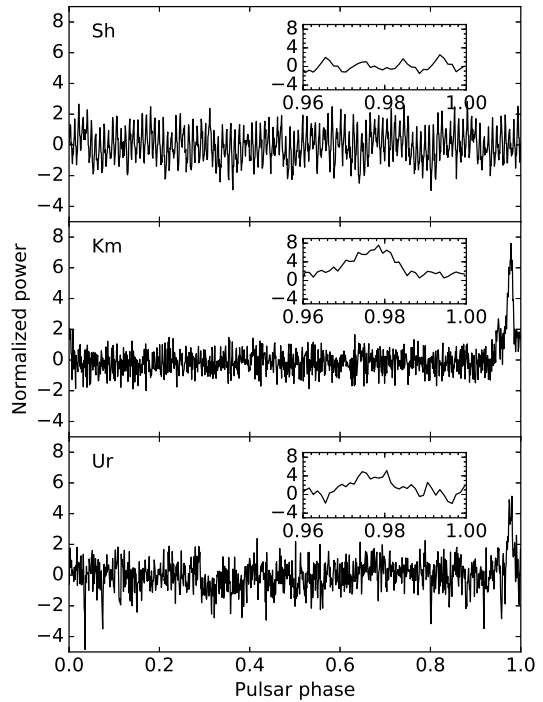}
	\caption{Pulse profiles of PSR J0334+5434. Derived by 
		folding the auto spectrum between 10~s and 170~s in pulsar observation 
		scan 
		73 for each station. The folding of scan 69 and 71 yield similar 
		profiles.  
		The small box inside each panel is the zoom in for profile between 
		pulsar phase 0.96 and 1.0. \label{fig:profile}}
\end{figure}

\subsection{Correlation and construction of time segments} 
\label{sec:corr_dedisp}
We use the CVN software correlator \citep{Zheng2010} for 
VLBI correlation. It was first developed for the Chinese Lunar Exploration 
Project (CLEP), and played important roles in the orbit determination of 
spacecrafts \citep{Zheng2007, Hu2008, Zheng2014}. 
Besides that, it has been used in the data processing of many geodetic VLBI 
observations \citep{Shu2009, CVN}. We have made extensive studies on 
the precision of its visibility output. Comparisons with DiFX
\footnote{
Available from the DiFX website at 
\href{https://www.atnf.csiro.au/vlbi/dokuwiki/doku.php/difx/start}
{\seqsplit{https://www.atnf.csiro.au/vlbi/dokuwiki/doku.php/difx/start}}}
 \citep{DiFX2007, 
DiFX2011} demonstrate that fringe fitting 
results of two correlators fit well. We use a modified version of the 
correlator for a better support of milliseconds visibility output. We choose 
PSR J0332+5434 in scan 69, 71 and 73 for single pulse 
search. The AP length is set to 1.024~ms. According to the observation 
schedule, the length of each scan is 180~s. 
Since the actual data starting and ending time of each scan are 
different for 3 stations, for consistency, we choose the data between 10~s and 
170~s of each scan and station for 
correlation and the subsequent single pulse search. The calibration source is 
3C273 in scan 293. Initial phase and channel delays are extracted from this 
scan using data between 10~s and 20~s. 
Time segments are constructed using the procedure in Sec. 
\ref{sec:time_seg}. According to Fig. \ref{fig:profile}, the pulse width is 
around 7~ms, which is 1\% of the pulsar period. As proposed in Sec. 
\ref{sec:method_cross_match}, we choose a window length of 4, 8, 16, 24 and 32 
APs.

\subsection{Pulse profile}
When processing VLBI pulsar observation data, we need to know the pulse profile 
of the corresponding pulsar, so as to set pulse gate to enhance SNR 
\citep{SFXC}. At present both DiFX and CVN software correlator are able to 
calculate pulse profiles by 
data folding. The idea is to first divide the pulsar phase from 0 to 1 into 
multiple phase bins, then place the auto spectrum of each frequency point into 
the corresponding phase according to the predicted pulsar phase. A pulse 
profile appears after enough time of folding. In this work, raw data recorded 
by each station are first time shifted according to VLBI delay models, such 
that all signals track the same wave front that passes through the geocenter. 
After that they are used for pulse profile folding and VLBI correlation. The 
pulsar phase prediction polynomials are generated with TEMPO2 for geocenter. 
Above operations guarantee the pulsar phase polynomials and detected signals 
are in the same geocentric reference frame, and therefore makes it possible to 
validate if a single pulse is pulsar signal or not with the predicted 
pulsar phase.

\begin{deluxetable*}{c|ccc|ccc|c}[t]
	\caption{Single pulse detection result with multiple baselines cross 
		matching scheme (Sec. \ref{sec:method_cross_match}). For each scan, we 
		list 
		the number of single pulses detected on at least 1, 2 and 3 baselines. 
		\label{tab:bl_cm}}
	\tablewidth{0pt}
	\tablehead{
		\colhead{Scan No.} & \multicolumn{3}{c}{1 baseline} & 
		\multicolumn{3}{c}{2 baselines} & \colhead{3 baselines} \\
		\colhead{}	& \colhead{Sh-Km} & \colhead{Sh-Ur} & \colhead{Km-Ur} & 
		\colhead{Sh-Km $\cap$ Sh-Ur} & \colhead{Sh-Km $\cap$ Km-Ur} & 
		\colhead{Sh-Ur $\cap$ km-Ur} & \colhead{Sh-Km $\cap$ Sh-Ur $\cap$ Km-Ur}
	}
	\decimalcolnumbers
	\startdata
	69	& 37 (12) & 33 (2) & 49 (40) & 1 (1) & 8 (8) & 1 (1) & 1 (1) \\
	71	& 26 (8)  & 35 (3) & 57 (41) & 1 (1) & 6 (6) & 1 (1) & 1 (1) \\
	73  & 29 (7)  & 34 (4) & 51 (36) & 2 (2) & 5 (5) & 3 (2) & 2 (2) \\
	\enddata
	\tablecomments{The number in the parentheses corresponds to single pulses 
		of which the pulse time range is overlapped with the pulsar phase 
		range.}
\end{deluxetable*}

\begin{figure*}[t]
	\plotone{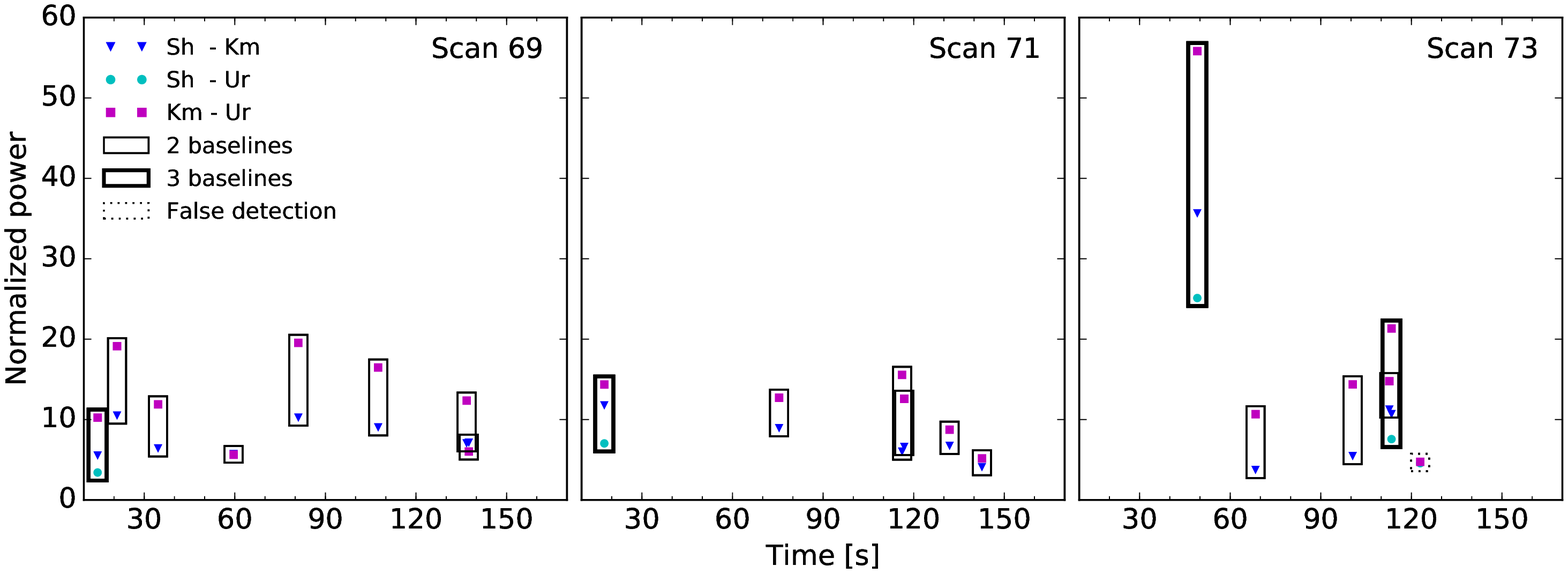}
	\caption{Single pulse detection result of multiple baseline cross matching 
		scheme in Sec. \ref{sec:method_cross_match}.
		Single pulses that belongs to the same cross matching group are 
		enclosed by 
		rectangular boxes. Thin and thick lines correspond to single pulses 
		detected on 2 and 3 baselines, respectively. Solid line suggests the 
		time 
		range of the single pulse is overlapped with the pulsar phase range. We 
		name them as ``high possibility pulsar signals''. The only false 
		detection 
		is enclosed by rectangular box with dotted line and locates at 122.9~s 
		in 
		scan 73. The corresponding pulsar phase is 0.255. Note that the actual 
		width of the 
		single pulse is much narrower than the width of the rectangular box. 
		The 
		detailed information of 4 single pulses detected on 3 baselines are 
		listed 
		in Tab. \ref{tab:result_3bl}. \label{fig:result_bl}}
\end{figure*}

\begin{deluxetable*}{cl|cccc}[t]
	\caption{Summary of single pulses detected on all 3 baselines. 
		\label{tab:result_3bl}}
	\tablewidth{0pt}
	\tablehead{
		\colhead{Scan No.} & \colhead{Baseline} & \colhead{Time} & 
		\colhead{Pulsar} & \colhead{{Width}} & \colhead{Normalized} \\
		\colhead{} & \colhead{} & \colhead{(s)} & \colhead{phase} &
		\colhead{(ms)} & \colhead{power}
	}
	\startdata
	& Sh-Km & 14.571136 & 0.978 & 24.576 & 5.530 \\
	69	& Sh-Ur & 14.562944 & 0.966 & 16.384 & 3.416 \\
	& Km-Ur & 14.567040 & 0.972 & 8.192 & 10.243 \\
	\hline
	& Sh-Km & 17.546880 & 0.975 & 4.096 & 11.761 \\
	71	& Sh-Ur & 17.548928 & 0.978 & 8.192 & 7.033 \\
	& Km-Ur & 17.546880 & 0.975 & 4.096 & 14.356 \\
	\hline
	& Sh-Km & 49.108608 & 0.976 & 4.096 & 35.642 \\
	73	& Sh-Ur & 49.108608 & 0.976 & 4.096 & 25.115 \\
	& Km-Ur & 49.108608 & 0.976 & 4.096 & 55.820 \\
	\hline
	& Sh-Km & 113.419904 & 0.974 & 8.192 & 10.667 \\
	73	& Sh-Ur & 113.417856 & 0.971 & 4.096 & 7.570 \\
	& Km-Ur & 113.417856 & 0.971 & 4.096 & 21.318 \\
	\enddata
\end{deluxetable*}

Fig. \ref{fig:profile} present pulse profiles of PSR J0332+5434 after 160 s 
folding. For 3 stations, Km exhibits the highest power, which is consistent 
with its antenna performance. Sh is strongly contaminated by RFI which makes it 
impossible to detect any meaningful signal 
in single dish mode. The profile of Km and Ur agree well with each other. we 
can make a raw estimation that the pulsar signal centers at 0.978 with a width 
of 0.01 in the pulse profile. In this work, a single pulse is assumed to be a 
``high possibility pulsar signal'' if its time range is overlapped with the 
pulsar phase range (0.973 - 0.983), since its possibility of pulsar 
origin is much higher than those outside of the pulsar phase range.

\subsection{Detection result}\label{sec:result_bl}
In this section, we present the single pulse detection result using two single 
pulse extraction schemes described Sec. \ref{sec:method_sp_extraction}, and 
give a brief comparison of two schemes.

\subsubsection{Multiple baselines cross matching scheme}
We set $N_\mathrm{win, min}$ as 3 for all 3 baselines, 
which means a single pulse is selected as a candidate signal (described in step 
d of Sec. \ref{sec:method_cross_match}) if it is detected in 3 or more windows. 
Then these candidate signals are cross matched on multiple baselines. 
We present the cross matching result of different baseline combinations 
in Tab. 
\ref{tab:bl_cm}. For single pulses that are detected on at least 1 baseline 
(column 2, 3, 4), a significant amount of them are those of which the time 
range is overlapped with the pulsar 
phase range (number given in the parenthesis of the table). Even for the Sh-Ur 
baseline of the lowest sensitivity (column 3), the fraction is 10\%, which is 
still tens times larger than the fraction of pulsar phase range among the
whole pulsar period (1\% according to Fig. \ref{fig:profile}). This excess can 
only be explained as single pulses that originate from the pulsar are 
detected with the cross spectrum method. Therefore, we name those single pulses 
as ``high possibility pulsar signals'' and define the detection confidence as 
the fraction 
of high possibility pulsar signals among all detected single pulses.
For 3 baselines, Km-Ur baseline yields the largest number of 
single pulses and the highest detection confidence, which is 
consistent with its high sensitivity. The low sensitivity of Sh 
antenna (Sheshan 25~m)
makes it difficult for single pulse detection on Sh related baselines, and 
further reduces the number of detected single pulses after cross matching.
All single pulses detected on at least 2 baselines are presented in Fig. 
\ref{fig:result_bl}. All of the 4 single pulses detected on 3 baselines are 
high possibility pulsar signals. For 16 single pulses detected on 2 
baselines (those detected on all 3 baselines are excluded), 15 of them 
are high possibility pulsar signals. Those results 
demonstrates that single pules detection on 2 or more baselines can almost 
exclude the possibility of false detection.
Note that the normalized power of some single pulses are actually  
small (less than 5). We detect them by first setting a low threshold 
to let signals in, then filtering RFIs with multiple windows.
 Also note that the power of detected single pulses varies one order 
of magnitude. This is consistent with the observed flux variation of PSR 
J0332+5434 in \citet{Kramer2003}.

\subsubsection{Multiple baselines power summation scheme}
To improve the detection efficiency, we carry out test of combining the power 
of several baselines before multiple windows filtering. The 
result of combining 2 and 3 baselines are presented in Tab. \ref{tab:bl_sum}. 
From the data, for the combination of Km-Ur baseline with one of the Sh 
related 
baselines (column 6, 7), the detection confidence is slightly enhanced. For the 
combination of all 3 baselines, the detection confidence does not increase. In 
both cases, fewer high possibility pulsar signals are detected compared with 
Km-Ur baseline only. One possible explanation is, we have fewer real pulsar 
signals that are detectable on low sensitivity baselines. Combining low 
sensitivity baseline with high sensitivity baseline means in most cases, we 
actually add noises to signals, and therefore reduce the SNR of high 
sensitivity baselines. Based on above analysis, it is not necessary to combine 
high sensitivity baseline with other baselines.
On the other side, for the 
combination of Sh-Km and Sh-Ur baselines (column 5), the detection confidence 
is 
greatly enhanced, although this is at the expense of losing 25\% of the high 
possibility pulsar signals. Besides that, compared with the same baseline 
combination of cross matching scheme, more high possibility pulsar signals 
are detected. Therefore, this scheme is indeed helpful for extracting weak 
signals from low sensitivity baselines.

As a summary, two cross spectrum based single pulse extraction schemes 
exhibit their own pros and cons. The multiple 
baseline cross matching scheme extracts single pulses with high confidence. 
However, only a small fraction of single pulses are detectable when the 
baseline sensitivity is low. The multiple baselines power summation scheme is 
able to extract more weak signals and enhance detection confidence on 
the combined low sensitivity baselines.

\begin{deluxetable*}{c|ccc|ccc|c}[t]
	\caption{Single pulse detection result with multiple baselines power 
		summation scheme (Sec. \ref{sec:method_power_sum}). For each scan, we 
		list 
		the number of single pulses detected on the combined baselines (column 
		5-8). For comparison, single pulses detected on every single baseline 
		are also listed (column 2-4), which are identical to corresponding 
		columns in Tab.\ref{tab:bl_cm}.\label{tab:bl_sum}}
	\tablewidth{0pt}
	\tablehead{
		\colhead{Scan No.} & \multicolumn{3}{c}{1 baseline} & 
		\multicolumn{3}{c}{2 baselines} & \colhead{3 baselines} \\
		\colhead{}	& \colhead{Sh-Km} & \colhead{Sh-Ur} & \colhead{Km-Ur} & 
		\colhead{Sh-Km + Sh-Ur} & \colhead{Sh-Km + Km-Ur} & \colhead{Sh-Ur + 
			km-Ur} & \colhead{Sh-Km + Sh-Ur + Km-Ur}		
	}
	\decimalcolnumbers
	\startdata
	69	& 37 (12) & 33 (2) & 49 (40) & 15 (9) & 42 (35) & 41 (33) & 40 (34) \\
	71	& 26 (8)  & 35 (3) & 57 (41) & 13 (5) & 39 (31) & 41 (32) & 30 (22) \\
	73  & 29 (7)  & 34 (4) & 51 (36) & 12 (6) & 34 (27) & 30 (28) & 21 (18) \\
	\enddata
	\tablecomments{The number in the parentheses corresponds to single pulses 
		of which the pulse time range is overlapped with the pulsar phase 
		range.}
\end{deluxetable*}

\subsection{Localization}
In this work, we did not carry out localization for the detected single pulses 
since it is almost impossible to obtain any reliable result by radio imaging 
with only 3 points in 
the UV plane. We have carried out study on the localization of point source 
using direct solving method, which behaves better under poor UV coverage. The 
idea is to express the differential phase as a linear function of 
position offset \citep{Duev2012}:
\begin{equation}
\Delta\phi = f (\Delta\alpha, \Delta\delta)
\end{equation}
where $\Delta\phi$ is the differential phase between the phase of target source
and the phase of reference source extrapolated to the target time, 
$\Delta\alpha$ and $\Delta\delta$ are the position offset to the a-priori 
position of the target source.
In theory, above equation is solvable with just two observables. However,
the solving process involves phase linking, ambiguity resolving, UV partials 
calculation, etc. We plan to cover them in our next work.

\section{FRB search with VGOS data}\label{sec:vgos}
The main motivation of this work is to investigate the possibility of FRB 
search in geodetic VLBI observations. In recent years, the International 
VLBI Service for Geodesy and Astrometry (IVS) proposed the 
VLBI2010 Global Observing System \citep[VGOS,][]{VGOS}, so as to fulfill the 
requirement of high 
precision measurement of geodesy and astrometry in the future. To achieve its 
goal, the whole system are updated, including small telescopes with 12-meter 
diameter, super wideband data recording system with a sample rate of 16-32 
Gbps, linear polarization, etc. On one hand, some features of the VGOS system 
make it very suitable for FRB search: large field of view (FoV), large number 
of stations, high frequency and long duration observation sessions. 
On the other hand, there are still some difficulties we must overcome. In this 
section, we investigate the possibility of carrying out FRB search in VGOS 
observation. 

\subsection{Detection sensitivity}
One disadvantage of FRB search with VGOS data is the size of the VGOS antenna 
is small, which leads to the relatively 
low sensitivity (SEFD $\sim$ 2000~Jy). However, this can 
be compensated with its large bandwidth ($\sim$~0.5~GHz). 
Based on Eq. \ref{eq:noise}, and take the SEFD listed in Tab. 
\ref{tab:antenna}, for a single pulse detected on Sh-Ur baseline with a window 
length of 4.096~ms and a total bandwidth of 96~MHz, the corresponding average 
noise level is 0.755~Jy. 
As demonstrated in panel b of Fig.~\ref{fig:fitdump}, the standard deviation 
(fluctuation) of noise is much smaller than the average noise level after 
fringe fitting. We may set 
the flux limit as this average noise level. Then a single pulse signal can be 
easily distinguished from noise fluctuations if its power is larger than this 
flux limit.
To achieve the same noise level for VGOS baselines, the corresponding 
window length for VGOS baseline is around 7~ms. Therefore, we may expect the 
sensitivity of VGOS baseline is sufficient for FRB search, if the flux of the 
single pulse is comparable with single pulses detected on Sh-Ur baseline, and 
the characteristic width is larger than 7~ms. Finally, we may derive a fluence 
limit of 5.3~Jy~ms for VGOS baseline.

\subsection{FFT size}
To fully exploit the relatively large FoV of VGOS antenna, we must estimate the 
appropriate FFT size when carrying out VLBI correlation for FRB search. In 
appendix \ref{append:FFT_size}, we have derived the relation 
of the minimum required FFT size with wavelength $\lambda$, antenna 
diameter $D$, baseline length $L_\mathrm{BL}$ and sampling frequency 
$f_\mathrm{s}$. According to VGOS specification, the 
512 MHz bandwidth of each band is divided into 16 frequency channels, which 
lead to a sample rate of 64 MHz for each frequency channel. Consider a typical 
VGOS baseline, $\lambda$ $\sim$ 0.136~m (S band, 2.2~GHz), $D$ $\sim$ 12~m, 
$L_\mathrm{BL}$ $\sim$ 3000 km (Shanghai to Urumqi), according to Eq. 
\ref{eq:FFT}, the minimum FFT size is 3551. We round it to 4096 
as a power of 2. The corresponding FFT length is 0.064~ms, which is much 
shorter than the typical FRB duration, and is therefore acceptable 
for VLBI correlation in FRB search.

\subsection{Detection rate}
\citet{Keane2015} report an FRB event rate of 2500 events per sky per day 
above a 1.4 GHz fluence of 2~Jy~ms. If we adopt an FRB spectral index 
$\gamma$ 
of 0.0 and a fluence index $\alpha$ of -1.0 as discussed in \citet{Caleb2017}, 
the FRB detection rate for VGOS baseline with a fluence limit of 5.3~Jy~ms at 
2.2 GHz is 0.0076 event per day. We have to point 
out that above estimation is conservative. First, the predicted FRB 
event rate is very uncertain. \citet{Champion2016} report the fluence-complete 
rate above 2~Jy~ms is $2.1^{+3.2}_{-1.5}\times10^3$ events per sky per day, 
which would double the VGOS detection rate if we adopt the upper limit.  
Besides that, above estimation assumes that antennas only observe one sky area 
in each scan. However, in real observation, antennas all over the world 
are usually subdivided into several groups, and observe two or three sky areas 
simultaneously. Taking these ingredients into account, we may expect an even 
higher detection rate. 

\subsection{Fringe fitting}
VGOS adopts super wideband data recording system and linear polarization, which 
requires improved fringe fitting algorithm to achieve its desired measurement 
precision. Some researchers already start their investigations, e.g, 
\citet{Cappallo2014, Kondo2016}.
Our next step is to combine the advantages of 
these algorithms and develop high performance fringe fitting algorithm for 
VGOS cross spectrum of milliseconds duration.

Based on above analysis, we may come to the conclusion that FRB search in VGOS 
observations is technically feasible. We do expect some exciting discoveries in 
the VGOS era, and will work on it. 


\section{Discussion}\label{sec:discuss}
\subsection{High performance implementation}
One disadvantage of the cross spectrum based search method is its high 
computation requirement due to the relatively complex algorithm: we have to 
carry out fringe fitting independently for a large number of time segments of 
different window lengths. However, some steps in the fringe fitting scheme can 
be optimized for a much better performance, which are introduced here.

One may find that the phase rotation by MBD and SBD in Eq. \ref{eq:fs_mod} 
are actually Fourier transforms, which means the searching of MBD and SBD that 
maximize delay resolution function can be implemented efficiently with a 2 
dimensional (2D) fast Fourier transform \citep[FFT,][]{NR}. In this case the 
delay 
resolution function becomes a matrix, with MBD and SBD take the gridded value 
$\Delta\tau_\mathrm{m,g}$ and $\Delta\tau_\mathrm{s,g}$, respectively. 
The process of maximizing delay resolution function becomes finding the maximum 
amplitude and the corresponding $\Delta\tau_\mathrm{m,g}$, 
$\Delta\tau_\mathrm{s,g}$ in the matrix. The detailed implementation 
and the discussion of ambiguities are available in \citet{Takahashi2000}. 

Moreover, we do not fit residual delay rate in the modified fringe fitting 
scheme. 
This leads to a good feature of Eq. \ref{eq:fs_mod}:
\begin{equation}
G_{k_0,~l_1} + G_{k_0 + l_1,~l_2} = G_{k_0,~l_1 + l_2},
\end{equation}
where $l_1$ and $l_2$ are window lengths. Above relation also holds for the 
delay resolution function matrix, which means the matrix of a time segment of  
large window can be expressed as the direct summation of matrices of several 
interconnected time segments of small windows. Mathematically above relation is 
trivial, however it greatly reduces the computation time of fringe fitting for 
multiple windows in real implementation.

Besides above two optimizations for the algorithm, almost every part of the 
fringe fitting scheme, including initial phase and channel delay calibration, 
2D FFT, amplitude calculation and maximum value localization of the delay 
resolution matrix, can be implemented in vectorized form. In fact, almost all 
modern numerical libraries provide support for these operations on 
different hardware platforms. It is not necessary to implement them from 
scratch. Moreover, there is no data exchange between time segments during 
fringe fitting. Therefore, the whole fringe fitting task can be easily divided 
into multiple small tasks based on time and baseline, and then assigned to 
multiple hardware processing units. One may fully utilize the computing power 
of model CPU/GPU clusters in this way.

\subsection{Difference with other cross spectrum based methods}
Our work is different from other cross spectrum based search methods when it 
comes to 
the detailed algorithm and implementation. In \citet{Marcote2017}, the arrival 
times of the bursts were first identified using Arecibo single-dish data. Then 
the pulse profiles were created with the total power of EVN cross correlations 
as a function of time. The exact time windows are determined in this way. After 
that the EVN data are dedispersed and correlated for these time windows. 
Finally the visibilities are calibrated and used for imaging of single pulses.
In \citet{Chatterjee2017}, two methods were used to detect bursts with 
fast-dump 
visibility data (milliseconds cross spectrum which is similar with time segment 
in our work) from VLA fast-dump observations. One was 
milliseconds imaging, which uses the realfast system to search burst in image 
domain \citep{realfast}. This 
method is based on the standard radio imaging pipeline, which involves phase 
calibration with reference source, 2D Fourier transform from UV plane to image 
plane, localization of single pulse, etc. \citep{Thompson2001}. Although 
realfast system 
already tries to simplify and automatize the whole pipeline, radio imaging is 
still a complex process and cannot be transplanted to other observation systems 
easily. Another method was beam-forming analysis, which carries out single 
pulse search on each synthesized beam. Similar method is used by CHIME, which
implements a real-time FFT-beamforming pipeline for FRB search \citep{CHIME}. 
 Our method can be regarded as an 
automatic version of the beam search algorithm. The fringe fitting process is 
somewhat similar with scanning the sky with a virtual beam. The power of 
the summed cross spectrum reaches maximum when the virtual beam points to 
the burst position. The difference is, there is no need to set the position of 
the beam in advance: the beam will find the burst position automatically via 
fringe fitting. 

\subsection{Search of dispersion measure}\label{sec:dm_search}
This work does not involve the search of dispersion measure. When constructing 
time segments in Sec. \ref{sec:corr_dedisp}, we just use the DM value in the 
reference. The reason is, the window length in the cross 
spectrum based method cannot be very 
small, otherwise the SNR is too low to detect any signal. As a 
result, this method is not sensitive to DM value. In other words, the DM 
resolution is low.  Here the DM resolution can be estimated with the minimum 
window length and the frequency range:
\begin{equation}\label{eq:dm_res}
\mathrm{DM}_\mathrm{res}=2.41\times 10^{-7}\times t_\mathrm{win,~min}~ /  ~
\left(\frac{1}{f_\mathrm{low}^{2}} - \frac{1}{f_\mathrm{high}^{2}}\right),
\end{equation}
where time and frequency take the unit of ms and MHz, respectively.
For a single pulse with DM value lower than DM resolution, the difference of 
arrival time within the frequency range is smaller than the window length. For 
the VLBI pulsar observation in Sec. \ref{sec:cross}, the minimum window length 
of 4.096~ms 
and a frequency range of 2192~MHz to 2288~MHz yield a DM resolution of 57.7 
$\mathrm{pc~cm}^{-3}$, 
which is not enough to resolve the DM value (26.833 $\mathrm{pc~cm}^{-3}$) of 
this pulsar. 

One possible solution for DM search in the cross spectrum based method is 
carried out in two steps: 
\begin{itemize}
\item[\textbullet] Coarse search. The DM search range is partitioned into 
multiple DM bins according to the DM resolution. For each DM bin, first 
carry 
out incoherent dedispersion and construct time segments (Sec. 
\ref{sec:time_seg}); then carry out fringe fitting for these time segments
(Sec. \ref{sec:fringe_fitting}) and extract single pulses (Sec. 
\ref{sec:method_sp_extraction}).
For a single pulse detected in several DM bins, pick up the bin that yields the 
highest signal power.
\item[\textbullet] Fine search. A DM search with higher DM resolution is 
carried out inside this DM bin. This can be achieved by either using the cross 
spectrum based method with higher resolution, or using the auto spectrum based 
method. The precondition for the latter choice is the signal is strong enough 
and therefore detectable in the auto spectrum of at least one station.
\end{itemize}
Consider a typical configuration of FRB search in VGOS observation, we set a 
minimum window length of 8 ms and a 0.5~GHz bandwidth (2.2~GHz $\sim$ 2.7~GHz). 
According to Eq. \ref{eq:dm_res}, the corresponding DM resolution is 27.8~pc 
~cm$^{-3}$, for the typical 2000 pc cm$^{-3}$ DM search range in FRB detection, 
this requires 72 bins in the coarse search step, which means we have to run 72 
times of the FRB search pipeline. As a first look, this requires a large amount 
of computation. However, as the most time consuming part of the whole pipeline, 
we already provide a set of high performance implementation for fringe fitting. 
In particular, our fringe fitting scheme is easy to be parallelized and 
accelerated with GPU. Therefore, technically, the large DM search range is not 
a big problem.

\section{Summary}\label{sec:sum}
In this paper, we introduce our cross spectrum based FRB search method. To test 
the method, we carry out single pulse search in a VLBI pulsar observation (3 
scans, 160~s per scan, 3 geodetic VLBI stations) and validate the result with 
predicted pulsar phase. The main points of this paper are summarized as follows.
\begin{itemize}
\item[\textbullet] The cross spectrum method takes the idea of geodetic VLBI 
data postprocessing and fully utilizes the fringe phase information to maximize 
the power of single pulse signals. To exclude RFI and achieve a higher 
detection confidence, candidate signals are filtered in multiple windows and 
cross matched on multiple baselines. Moreover, the power of multiple 
baselines are combined to improve detection sensitivity.

\item[\textbullet] According to our test, single pulses detected on 2 or 
more 
baselines can almost exclude the possibility of RFIs. By combining the power 
of multiple baselines, a greater number of weak signals are extracted from low 
sensitivity baselines with higher confidence. 

\item[\textbullet] The whole pipeline is easy to implement and parallelize, 
which can be deployed in various kinds of VLBI observations. In 
particular, we point out VGOS observations are very suitable for FRB search.
\end{itemize}

\acknowledgments
The authors thank Wu Jiang for kindly providing VLBI pulsar observation data
and appreciate the support of CVN data processing center. The authors thank
Aard Keimpema, Roger Cappallo, George Hobbs, Tetsuro Kondo, Jintao Luo, Yajun 
Wu, Zhen Yan, Maoli Ma, Wentao Luo for helpful discussions. This work is 
sponsored by the Natural Science Foundation of China (11373061, 11573057), the 
Key Laboratory of Radio Astronomy of Chinese Academy of Sciences, the CAS Key 
Technology Talent Program, 
the Science and technology infrastructure platform project of National Science 
and Technology Ministry ``National Basic Science Data Sharing Service 
Platform'' 
(DKA2017-12-02-XX),
the Chinese Academy of Sciences innovation project 
(CXJJ-17-Q113), the Chinese lunar exploration project, the Natural Science 
Foundation of Shanghai (15ZR1446800).
\software{
PRESTO \href{http://www.cv.nrao.edu/~sransom/presto}
	{\seqsplit{(http://www.cv.nrao.edu/\textasciitilde{}sransom/presto)}},
HOPS \href{http://www.haystack.edu/tech/vlbi/hops.html}
	{\seqsplit{(http://www.haystack.edu/tech/vlbi/hops.html)}},
DiFX \href{https://www.atnf.csiro.au/vlbi/dokuwiki/doku.php/difx/start}
	{\seqsplit{(https://www.atnf.csiro.au/vlbi/dokuwiki/doku.php/difx/start)}}
}

\appendix
\section{Initial phase and channel delay extraction}
\label{append:pcal}
The extraction of initial phase and channel delay can be summarized in the 
following steps. 
\begin{itemize}
\item[a] Setting $\phi_0^n$ and $\Delta\tau_0^n$ as 0. 
\item[b] Constructing a time segment with a window length of several seconds 
for the calibration source\footnote{This is based on the assumption that the 
clock has been well 
adjusted and the residual fringe rate for the correlation result of the 
calibration source is small. E.g., less than $10^{-3}$ Hz. Therefore the cross 
spectrum can be integrated for a longer time. If this is not the case, we have 
to use the standard fringe fitting scheme which fits residual delay rate as 
well.}. 
\item[c] Carrying out fringe fitting on this time segment and finding out 
$\MBD$ and $\SBD$ that maximize delay resolution function. The initial phase of 
frequency channel $n$ is derived by:
	\begin{equation}
	\phi_0^n	=	\angle\sum_{j=1}^{J-1}S_{k_0,~l}(n,j)e^{-i2\pi\left[f_j\SBD 
		+ (f_0^n - f_\mathrm{ref,fit})\MBD\right]}.
	\end{equation}
The channel delay $\Delta\tau_0^n$ is derived by fitting 
	$S_{k_0,~l}(n,j)e^{-i2\pi f_j\SBD}$. 
\end{itemize}
The initial phase extraction method introduced here is similar with the 
``manual'' mode in HOPS. One may use the PCAL signal extracted by HOPS 
directly. 

\section{Estimation of FFT size}
\label{append:FFT_size}
Since delay models are calculated for the center of the FoV, 
a source that does not appear in the center of FoV will lead to an extra 
delay $\Delta\tau$, which varies from $-\Delta\tau_\mathrm{max}$ to 
$\Delta\tau_\mathrm{max}$. The maximum delay $\Delta\tau_\mathrm{max}$ is 
achieved when the source appears at the edge of the main beam. The size of the 
main beam (angular diameter) can be estimated with the full width at half 
maximum: $\theta_\mathrm{FWHM}\sim\lambda/D$. Here $\lambda$ is the wavelength, 
$D$ is the antenna diameter. Then $\Delta\tau_\mathrm{max}$ can be estimated 
with: 
$\Delta\tau_\mathrm{max}\sim\frac{1}{2}\theta_\mathrm{FWHM}~L_\mathrm{BL}/c$. 
Here $L_\mathrm{BL}$ is the baseline length.
 
In the FX type correlator, two time series are first Fourier transformed to 
frequency domain via FFT with size $N$, the corresponding 
frequency points are conjugate multiplied. The cross spectrum is obtained in 
this way with frequency resolution $f_\mathrm{s} / N$. Here $f_\mathrm{s}$ is 
the sampling rate of each frequency channel. 
In general, for each frequency channel, the cross spectrum with 
frequency resolution $f_\mathrm{s} / N$ corresponds to the SBD search range 
from $-\frac{N}{2}~T_\mathrm{s}$ to $\frac{N}{2}~T_\mathrm{s}$. Here 
$T_\mathrm{s}$ is the sampling time: $T_\mathrm{s}=1/f_\mathrm{s}$. When 
searching for SBD in the real fringe fitting scheme, the cross spectrum will be 
zero padded to achieve a higher time resolution. 

To make the source detectable, the SBD search range in the fringe fitting 
scheme must cover the maximum delay: $\frac{N}{2}T_\mathrm{s} \ge 
\Delta\tau_\mathrm{max}$. Finally we get the minimum required FFT size:
\begin{equation}\label{eq:FFT}
N_\mathrm{min} = \frac{\lambda}{D}\frac{L_\mathrm{BL}}{c}f_\mathrm{s}.
\end{equation}

\end{document}